# Complex-network approach for visualizing and quantifying the evolution of a scientific topic

Olesya Mryglod[1,4], Bertrand Berche[2,4], Yurij Holovatch[1,4], Ralph Kenna[3,4]

[1]*Institute for Condensed Matter Physics of the National Academy of Sciences of Ukraine, 79011 Lviv, Ukraine*
[2]*Université de Lorraine, Laboratoire de Physique et Chimie Théoriques, 54506 Vandœuvre lès Nancy Cedex, France*
[3]*Applied Mathematics Research Centre, Coventry University, Coventry, CV1 5FB, England*
[4]$\mathbb{L}^4$*Collaboration & Doctoral College for the Statistical Physics of Complex Systems, Leipzig-Lorraine-Lviv-Coventry, Europe*

**ABSTRACT**

*Tracing the evolution of specific topics is a subject area which belongs to the general problem of mapping the structure of scientific knowledge. Often bibliometric data bases are used to study the history of scientific topic evolution from its appearance to its extinction or merger with other topics. In this chapter the authors present an analysis of the academic response to the disaster that occurred in 1986 in Chornobyl (Chernobyl), Ukraine, considered as one of the most devastating nuclear power plant accidents in history. Using a bibliographic database the distributions of Chornobyl-related papers in different scientific fields are analysed, as are their growth rates and properties of co-authorship networks. Elements of descriptive statistics and tools of complex-network theory are used to highlight interdisciplinary as well as international effects. In particular, tools of complex-network science enable information visualization complemented by further quantitative analysis. A further goal of the chapter is to provide a simple pedagogical introduction to the application of complex-network analysis for visual data representation and interdisciplinary communication.*

**INTRODUCTION**

The subject of this chapter unites two major topics: complex-network science and scientometrics. Both topics reflect new trends in the evolution of science and, on a more general scale, in the evolution of human culture as a whole. Both are tightly related to the notion of complex systems which is gradually appearing as one of central concepts of our times. Usually by "complex system" one means a system that is composed of many interacting parts, often called agents, which display collective behavior that does not follow trivially from the behaviors of the individual parts (for recent discussions see e.g. Thurner 2017, Holovatch et al. 2017). Inherent features of complex systems include self-organization, emergence of new functionalities, extreme sensitivity to small variations in initial conditions, and governing power laws (fat-tail behaviour). In this sense science itself, both as an enterprise that produces and systematizes knowledge as well as the body of this knowledge, is an example of a complex system. The above inherent features are ubiquitously present in different forms of scientific activity. Some of them will be a subject of analysis in this chapter.

In the field of scientometrics, science itself is the subject of analysis: What is the structure of science? How do interactions between different scientific fields occur? What is the impact of a certain field, certain scientific article or of certain academic institutions? These and many more questions lie within the parameters of scientometric studies. Numerous indicators have been invented in order to quantify answers to these questions. Currently almost everyone in the academic world is familiar with metrics such as the impact factor, citation index, Hirsch factor and similar 'magic' numbers frequently used in science-policy and management contexts. These metrics are usually based on citation and bibliographic data. To acquire interpretable information from such large amounts of interconnected data, special tools must be used. The results are highly applicable in this case; new knowledge about how the system of science is organized can be used to improve it on different scales. Therefore visual representation of data and of the results of its analysis becomes vital.

An analysis of the evolution of a scientific topic is considered in this chapter. Such a case study is used here to describe some of the tools and methods useful for bibliographic data analysis. The questions discussed in our work (Mryglod et al. 2016) concern a very subtle effect one can experience in the evolution of scientific studies: How does a new topic in science emerge? When does it appear? How does it evolve? What are the principal factors involved in its development? Amongst different ways to approach these questions, and to quantify the answers, an obvious idea is to analyze the dynamics of scientific publications paying attention to content analysis, disciplinary targeting, institutional involvement, etc. In the case study presented below, we analyze the academic response to the disaster that occurred in 1986 in Chornobyl (Chernobyl), Ukraine, considered as one of the most devastating nuclear power plant accidents in history (Alexievich & Gessen 2006).

The science of complex networks gives an essential methodological framework enabling one to describe complex-systems behaviour in a quantitative and predictive way (Albert & Barabási 2002, Dorogovtsev & Mendes 2003, Newman, Barabási, & Watts 2006). Moreover, another goal of a complex-network approach is that very often it allows one to visualize the system under consideration. In turn, both quantitative descriptions and visualizations enable understanding of complex-system behaviour. One of the goals of our paper is to show how such an approach works when combined with other ways of scientometric analysis using the above case study as an example.

The rest of this chapter is organized as follows: first we make a very short introduction to the field of complex networks, giving main definitions and introducing principal observables used to quantify them. We then we pass to the above question of interest, displaying some of our results that concern quantification of the emergence and evolution of a scientific topic. There, we use complex-network representations to visualize certain processes involved and complement their analysis by elements of descriptive statistics. We conclude by summarizing results obtained and giving an outlook.

## COMPLEX NETWORKS: BACKGROUND

Here, we provide a brief introduction of what network science is about and define some of the main observables used to quantify complex networks.

A *network* is an assemblage of *nodes* connected by *links* (also called edges), see Figure 1. In our case the nodes will be countries and links between pairs of them are formed if authors from each have jointly published a paper on the Chornobyl topic, which is recorded in the *Scopus* database. The degree of a node is the number of links emanating from it. In principle, if there are $N$ nodes in the network, they could be interconnected by $N(N-1)/2$ links. If the actual number of links in a network is $M$ instead, the density of the network is $2M/[N(N-1)]$.

Some parts of a network may be denser than others and there are various ways to capture such features. Indeed, the network itself may be fragmented into a number of *components* which are not connected to each other through links. The largest possible component, in principle, is the entire network itself (if each node can be reached from every other node by a sequence of links). The smallest possible component is a single isolated node, disconnected from all other nodes of the network. Usually the largest component is smaller than the entire network and in this case it is called the giant component. This is usually our focus of interest in such fragmented networks.

We are frequently interested in statistics involving lengths of paths in the network. Paths between nodes are formed by sequences of links. The shortest path between any two nodes is called a *geodesic*. If the network is fragmented, we only consider geodesics within a given fragment because there are no paths interconnecting two separate fragments (in a sense the distance between them is infinite). The longest finite geodesic in the entire network is called the network *diameter*. This is the longest shortest path in the network. We are also interested in the *mean path length* formed by averaging over the distances between all pairs of nodes which are connected by a path. This notion became famous by American psychologist Stanley Milgram's experiment in the 1960s to investigate the "small world problem" (Milgram 1967). This was related to a suggestion that, despite the world's population of over 7 billion people, everyone is connected to everyone else by six or fewer steps or "a friend of a friend" statements.

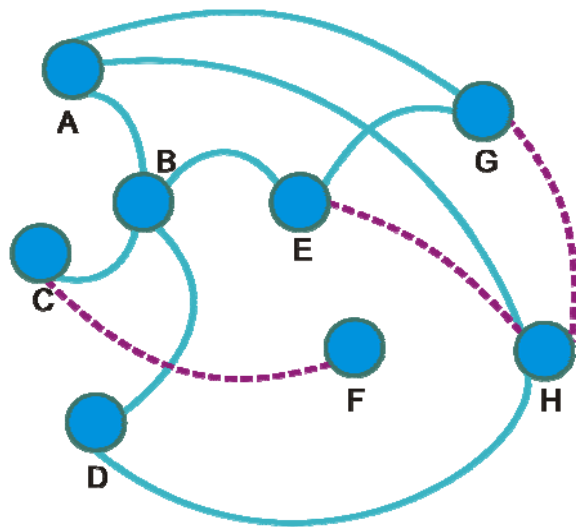

**Figure 1:** (Color online) A small network comprising $N = 8$ nodes interconnected by $M = 11$ links. The nodes may represent people in real life, characters in a text, stations in a transport system or countries producing scientific publications. Links represent some sort of interactions or relationships between them. In network science one seeks to capture the statistics relating to how these links are distributed. In social networks links may have different features, e.g., here, the solid light-blue links and the dashed purple ones may represent positive (friendly) and negative (hostile) relations.

In Figure 1 we present a simple network for demonstration purposes. The network has $N = 8$ nodes interconnected by $M = 11$ links. Since 8 nodes could, in theory, be connected together by $N(N-1)/2 = 28$ links, the density of the network is 11/28 = 0.39. The degree of each node varies, e.g., Node A has degree 3 while B has four links. The average number of links per node is 11/8 = 1.375. There is no fragmentation in this network – every node belongs to the giant component. The network diameter is 4 because the longest shortest paths (longest geodesics) have this length. These are paths joining F to G (namely FCBAG or FCBEG) or F to H (FCBEH).

The *clustering coefficient* provides one measure of how dense a local part of a network is. In a social network, for example, if an individual has two friends, it is fairly likely that these two are also acquainted. In Figure 1 node A has 3 neighbours (namely B, G and H). There are therefore three potential relationship triads involving node A, (namely ABG, ABH and AGH). Of these, only one is realised (namely AGH). We say that the clustering coefficient for node A is one out of three or 1/3. We may then proceed and calculate the clustering coefficients for each node on the entire network. Taking an average of all of the resulting numbers is gives the clustering of the entire network.

If a network has a small path length and a high clustering coefficient, we say that it is a *small world*. To decide how small the path length has to be and how high the clustering has to be for a network to be small world, we compare to a random graph of the same size and average degree. If the path lengths are comparable in magnitude but the clustering of the network at hand is far greater than that of the random network, the network is deemed small world. In small worlds, every node is close to every other node, in a sense.

Sometimes a subset of nodes can be very highly clustered – very tightly connected. If a subset of nodes is complete, i.e., if every distance pair of nodes is connected by a link, it is called a *clique*.

It is often desirable to decide if some nodes are particularly important or influential. One way to do this is to consider that those with highest degree are most important. Another is to consider nodes with the highest *betweenness* centralities as most important. Betweenness counts the number of shortest paths (geodesics) which pass through the given node. To define it for a specific node, we count all geodesics that go through it and divide that by the total number of paths on the network. We then divide the result by $(N-1)(N-2)/2$ in order to normalise it in such a way that it is one if all geodesics pass through node.

Another measure of the importance of a node is its *closeness* centrality. This is defined by first taking the sum of the distances from the given node to all other nodes in a connected component of the network. This is termed its farness. The reciprocal of farness is a simple measure of how central a node is and is termed its closeness.

Many other statistics have been invented to capture and compare various characteristics of networks. In particular, defining the mean degree one naturally arrives to the notion of node *degree distribution*: a probability to find a node of given degree. The form of such distribution is crucial in defining network properties. For many complex networks it decays as a power law. Such networks are called scale-free. But here we have given the measures that are used in the remainder of this chapter. The reader is referred to the literature (e.g., Newman 2010) for more extensive discussions of multitudes of such measures.

**VIZUALIZING THEMATIC COAUTHORSHIP: A CASE STUDY**

Complex-network theory provides a set of useful tools to analyze also bibliographic data. For this reason modern scientometrics operates with different kinds of networks such as citation and co-citation networks, coauthorship networks or those which represent scientific papers connected by common authors or key words. In particular, coauthorship data constitute one of the most useful sources of information for detection of links between papers, journals or even countries. To give an example of such an analysis we refer to our recent work where a case study is considered (Mryglod 2016). These results contribute to a direction of scientometrics research focused on the analysis of the evolution of separate scientific topics. The knowledge about how new topics emerge and evolve is useful, in particular, to describe the forefront of research and to reveal the structure of scientific disciplines as well as their interconnections.

The main aim of our work was to study the reaction of academic community to a particular important event. Similar to how society's response can be tracked through the analysis of news, discussions in mass media or activity in social networks, the interest of academics is reflected in scientific publications. Since periodicals provide the most common way to present and interchange new results, it is natural to consider the set of publications in scientific journals. To give an example, the bibliographic data about papers relevant to the problem of poverty were analyzed in (Zuccala, A. & van Eck 2011). General publication activity on a given topic, interdisciplinary landscape, as well as international collaboration patters, was studied in this work. We applied the suggested set of methods, not to analyze the core of papers relevant to a permanently important problem, but rather to highlight the emergence of new trends in response to single event.

The Chornobyl catastrophe was chosen as the trigger event for our study. Several reasons (besides the fact that two of us are Ukrainians) led to such a choice:
- The consequences of this disaster can be felt in different spheres of humanity and, therefore, attract the attention of researchers of various disciplines.
- The worldwide scale of the Chornobyl accident implies international interest in the problem. At the same time the reaction of the academic community on a national level is strong enough to make comparisons and to reveal the role of Ukraine in the international scene within the topic.
- The exact time of "birth" can be defined for Chornobyl-related research topics: 26 April 1986 (of course, papers using name Chornobyl and published before 1986 can also be found but the number of post-accident paper is obviously larger: 1.8 publication per year on average between 1966 and 1985 comparing to above 200 publications already in 1986).

Over 9500 records about research papers containing words "Chornobyl" (and its variations: "Chornobyl'", "Chernobyl" or "Chernobyl'") in their titles, abstracts or authors' key words were collected from Scopus[1]. Almost 75% of records contain the "affiliation" field, which is useful to find countries related to authors. Using these data international collaborations on the Chornobyl topic are studied. The coauthorship network at the level of countries is built; nodes represent separate countries while each link connects two nodes if two different countries were mentioned in the affiliation list of the same publication. The aggregated network incorporating all the data starting from 1986 and until the beginning of 2015 is considered below. Therefore, it also includes a number of countries that do not exist anymore as political entities, such as Czechoslovakia or the USSR.

The network consists of 97 nodes connected by 761 links (see Table 1 for these and some other numerical characteristics of the network). On one hand, such a ratio indicates quite a sparse network; only approximately 16% of possible links are actualized. On the other hand, most nodes are mutually reachable since over 82% of them belong to the largest connected (giant) component. This means that the majority of countries are interconnected through Chornobyl-related research collaboration. Moreover, starting from a randomly chosen country, one can reach another one in just 2 steps on average. The longest geodesic (network diameter) connects Yugoslavia and the Faroe Islands and still takes only 4 steps. It is interesting to note that the large connected component is in fact the only connected fragment in the network. The remaining 17 nodes are isolated: the authors from countries such as Lebanon, Iran, Pakistan and others were not involved in the international collaboration on Chornobyl topic.

---

[1] The data collected during January–February 2015 are used.

The remaining network statistics are given in Table 1 can be easily interpreted as well. The average value of the node degree indicates that each country collaborates with more than 15 other countries on average. The highest level of collaboration activity, represented by the maximal node degree, corresponds to the USA: 1082 papers (the second largest number after Russia which produced 1180 publications) were coauthored with 51 other countries. Being the largest hub, the USA also occupies an influential place in the network. This is confirmed by special quantitative node properties – betweenness centrality and closeness centrality. The maximal value of the former means that the largest number of shortest paths between any other two countries passes through the USA. Taking into account that such shortest paths on average consists of 2 steps, the USA appears to be a direct bridge between the majority of the other countries. The latter centrality value gives the closeness of a particular node to any other node in the network – and it is maximal for USA as well.

The Watts-Strogatz clustering coefficient is another numerical value which describes the network connectedness. It reflects the average probability of nodes to be a part of the fully connected clusters – *cliques*. Clustering coefficients can be calculated for connected fragments that imply that the majority of nodes are taken into account in our case. The neighbours of a node are interconnected in their own turn with probability 0.78.

**Table 1**. Some of numerical values which characterize coauthorship network on the level of countries for Chornobyl-related publications (1986 – January 2015) visible in Scopus.

| Number of nodes | Number of links (network density) | Average node degree | Max degree | The longest geodesic | Watts-Strogatz clustering coefficient | Number of isolated nodes (%) | Number of nodes within the largest connected component (%) |
|---|---|---|---|---|---|---|---|
| 97 | 761 (0.16) | 15.7 | 51 | 4 | 0.78 | 17 (17.5%) | 80 (82.5%) |

A visual representation of the network (Figure 2) provides a more intuitive way to understand the structure discussed above. One can instantly distinguish the connected component surrounded by a comparatively small set of isolated nodes. Textual labels with short names of countries can also contain the exact values of node attributes: total numbers of papers, population, degree values or any others. However, it is often hard to visualize large network consisting of a lot of nodes and numerous links in a simple and intuitively clear way. Therefore, it is convenient to use different possible "dimensions" of visual attributing of network representation depending on the purpose. The layout, size and color can be used to highlight particular fragments of the network or to distinguish nodes or links by an attribute.

The same coauthorship network as in Figure 2 is presented in Figure 3 using a different kind of representation:
- TOP10 nodes characterized by highest degree value are positioned in the center;
- The links connecting central nodes are highlighted in yellow;
- Node size is proportional to the total number of papers, where a particular country is mentioned in affiliation list.

Such a visualization approach allows one instantly to see the most active countries – collaborators within the Chornobyl-related topic – and to compare the absolute contributions in terms of numbers of publications. One can wonder about the factors which play a role to manifest these hubs: geographical closeness to the epicenter of the tragedy (Ukraine, Russian and Belarus) and the active position of developed countries (USA, United Kingdom, France, Germany, Austria, Japan, Denmark). Eight of these countries are also in the TOP10 list of the most productive countries in terms of numbers of publications (Austria and Denmark were in 12th and 21th positions, respectively, in the most recent rating).

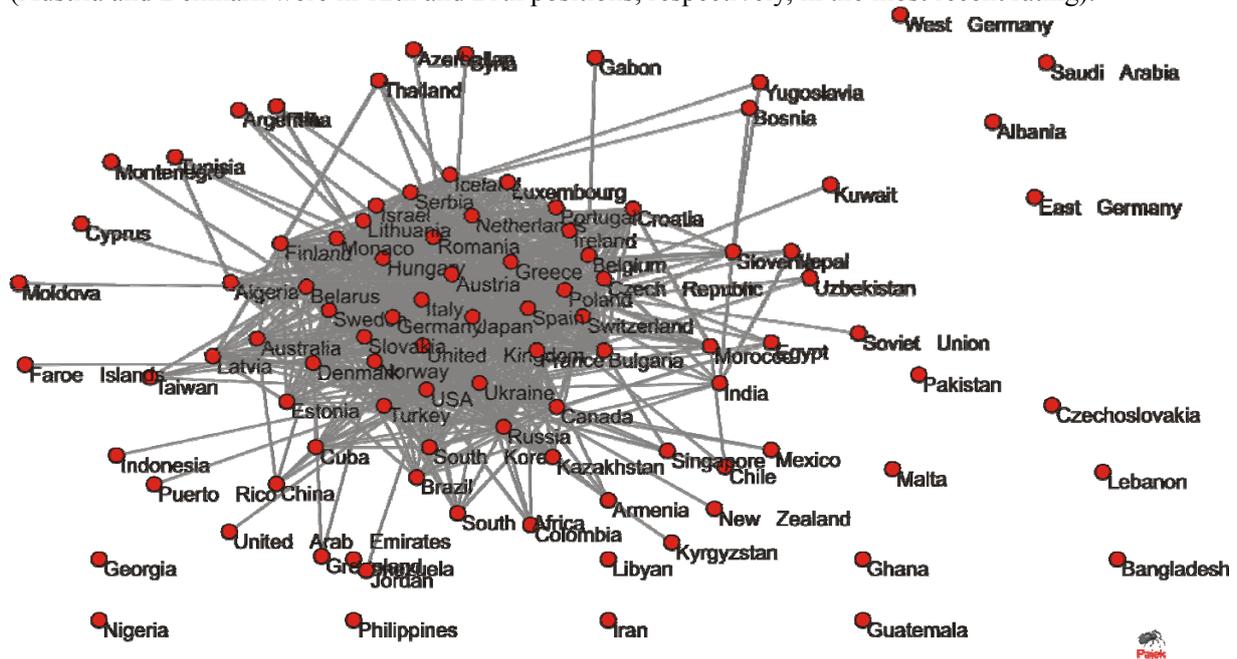

**Figure 2**. (Color online) Collaboration network of countries for Chornobyl-related research based on Scopus data (publication period: 1986–2015). Visualization of this and further networks is performed using Pajek network visualization software (Vlado 2015).

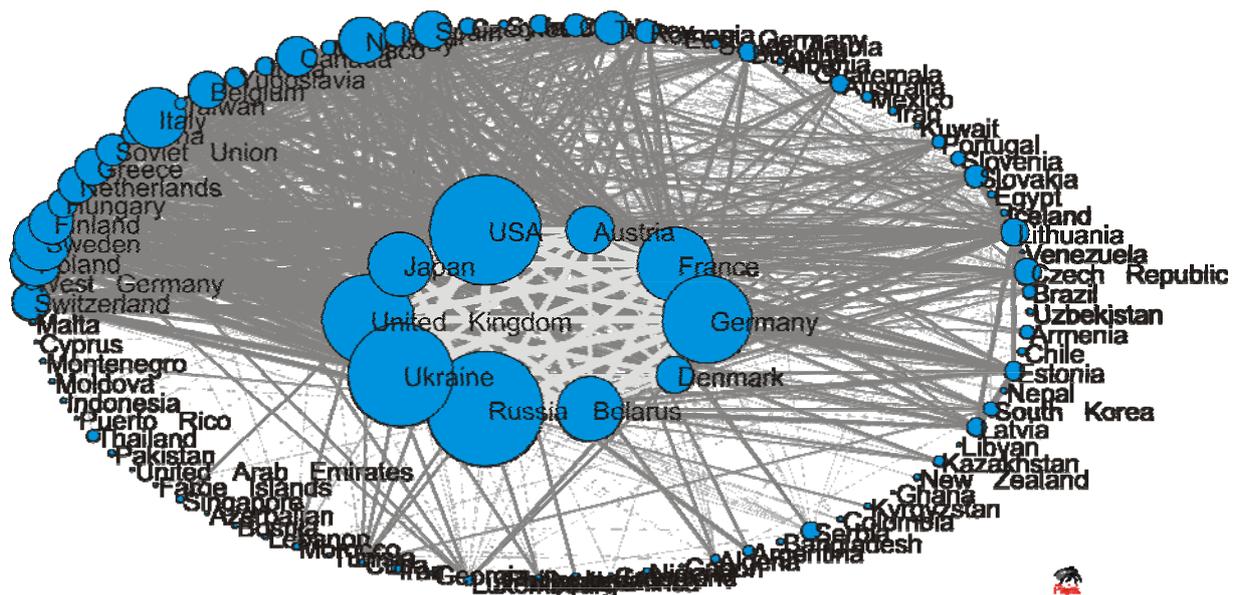

**Figure 3.** (Color online) Collaboration network of countries for Chornobyl-related research based on Scopus data (publication period: 1986–2015). TOP10 of nodes with the highest degrees are centered and the links connecting these nodes are highlighted in lighter color. Node size is nonlinearly proportional to the total number of papers, published by authors from the corresponding country. Link widths are nonlinearly proportional to the numbers of common publications by authors from two given countries.

The highly connected fragment highlighted in yellow in Figure 3 demonstrates the high level of connectedness amongst the TOP10 most actively collaborating countries which is rather natural. In fact, these countries form a clique.

Another way of representing of the same coauthorship network is used in Figure 4:
- The nodes colors and the layout depend on the part of the world each country belongs to;
- Node size is proportional to the node degree.

In this case one can explore the geographical aspect of international collaboration. The differentiation of the nodes by their degree allows one to speculate about the typical collaboration patters. E.g., it is easy to distinguish the most connected countries for the Asia region: Russia and Japan each have 42 coauthorship links to other countries while a majority of the rest countries has one or no links. A more balanced situation is observed for European countries: e.g., 8 of them have at least 40 collaborators and 9 of them have no more than 1 collaborator.

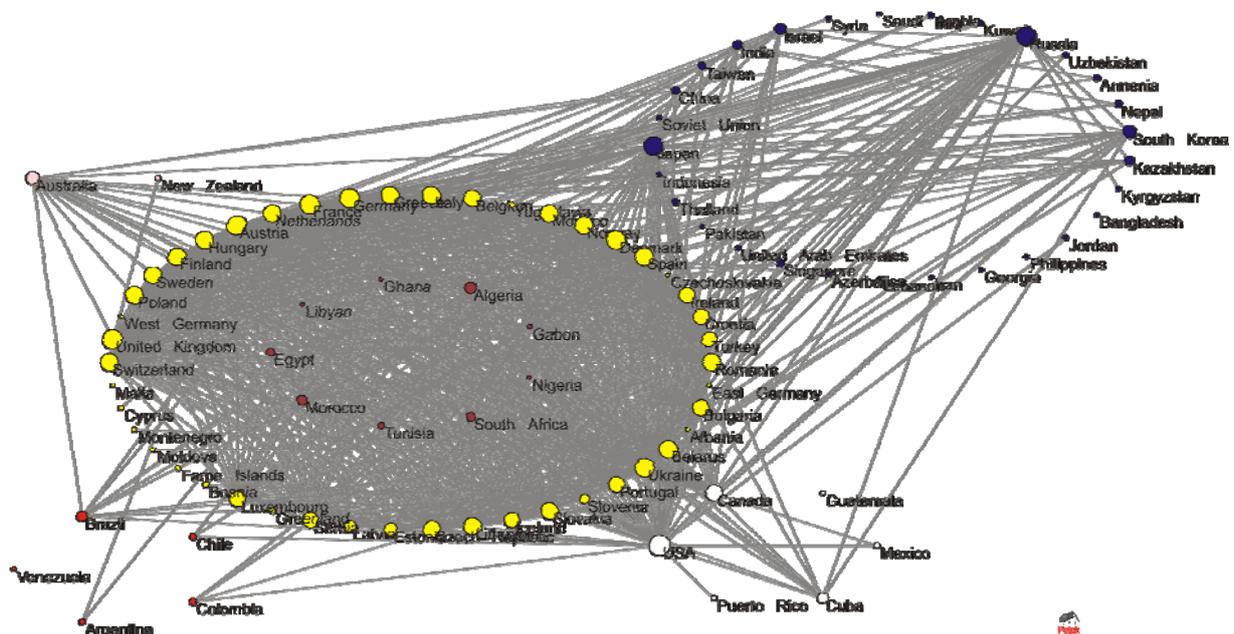

**Figure 4.** (Color online) Collaboration network of countries for Chornobyl-related research based on Scopus data (publication period: 1986–2015). The nodes of different kinds (different colors online) represent different part of the world – for the sake of visual simplicity the same attribute is used to include a node to corresponding circle. A node size is proportional to its degree.

Finally information about the year when each country introduced the first Chornobyl-related publications is used in Figure 5. Again, we present the same network, but this time visualized in such a way to see how countries from different parts of the world entered "into the game". Obviously, the largest number of countries reacted immediately after the accident in 1986 or one year later. A sort of anniversary effect can be seen here: new countries joined 5 (6) and 10 years later. It is natural that geographical position is important: the first reaction wave came mostly from Eurasian countries. Supposedly, the more distant countries started to discuss the Chornobyl problem (i) when remote consequences became worldwide and (ii) after Fukushima catastrophe in 2011.

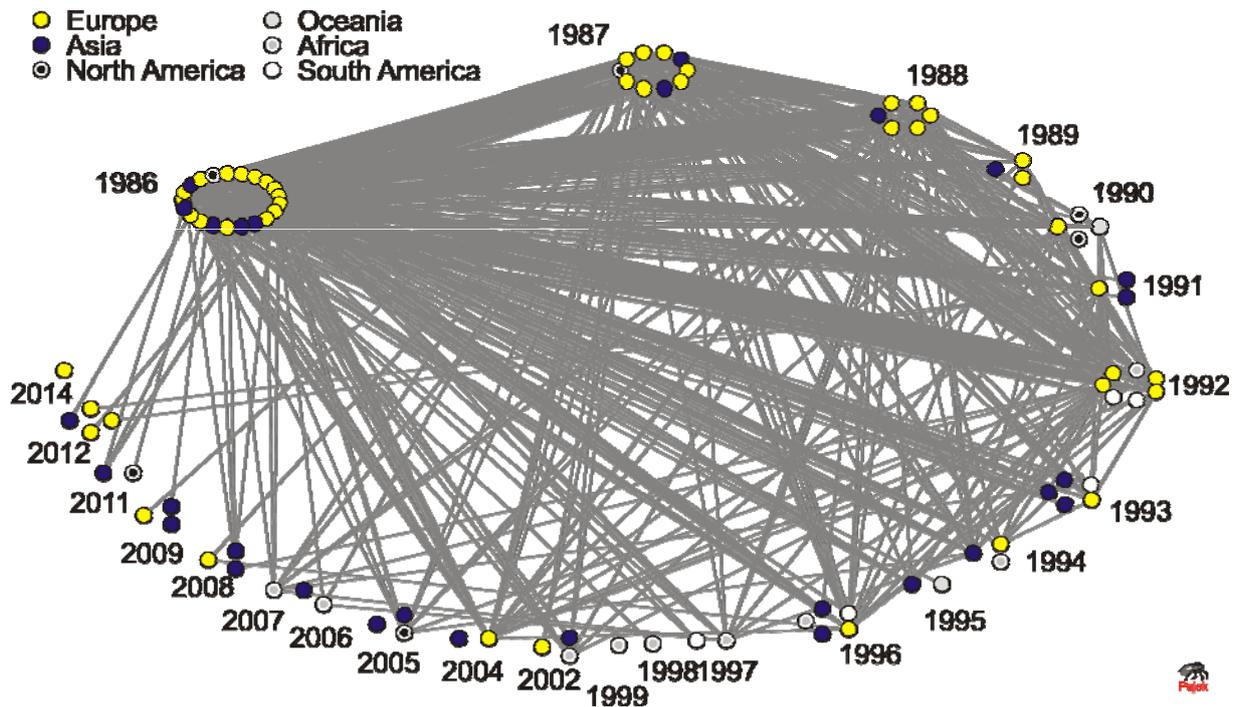

**Figure 5.** (Color online) Collaboration network of countries for Chornobyl-related research based on Scopus data (publication period: 1986–2015). The nodes of different kinds (different colors online) represent different part of the world. Nodes are grouped by the year of the first Chornobyl-related publication of a country.

Besides the variety of ways to represent the same data, network theory allows us to consider "slices" of network. Any node attributes can be used to group them and to extract subnetworks for further consideration. In Figure 6 the evolution of the network described above can be seen: the first subfigure represent the picture of collaboration during the first several years after the accident (1986–1990), the second subfigure shows the network built on the data about publications during the several later years (2011–2015). The so-called sliding window technique is preferable due to a change of historical circumstances. This means that different periods of time are considered and separate coauthorship networks are built for each of them taking into account only data "visible" within the particular time window. Such an approach is also useful in order to see the non-aggregative nature of collaboration patterns dynamics: new links can appear while some old relations become broken. Thus, the visual densification of the network can be clearly seen in Figure 6. Such an effect when the network becomes denser accumulating more nodes and links was described for different real networks (Leskovec et al. 2007). To check it numerically for our network its evolution is considered as an integral process, i.e. all the links and nodes are taken into account starting from 1986. Similarly as in (Leskovec et al. 2007) the increase of the average node degree is observed: it grows starting from 1.4 for the first snapshot made for interval 1986-1990 and ending with 15.7 for 1986-2015 data. The network diameter decrease is not observed supposedly due to the small statistics, but this value is obviously not increasing – it fluctuates around 4. The number of links increases much faster than the number of nodes: it is also hard to make a reliable quantitative conclusion about the shape of this dependence, but it is not far from being power-law (see Figure 7) which characterizes the network densification process. Such a densification can be interpreted as the intensification of international collaboration about the topic: along with new countries joining new links between the former members of collaboration network appears.

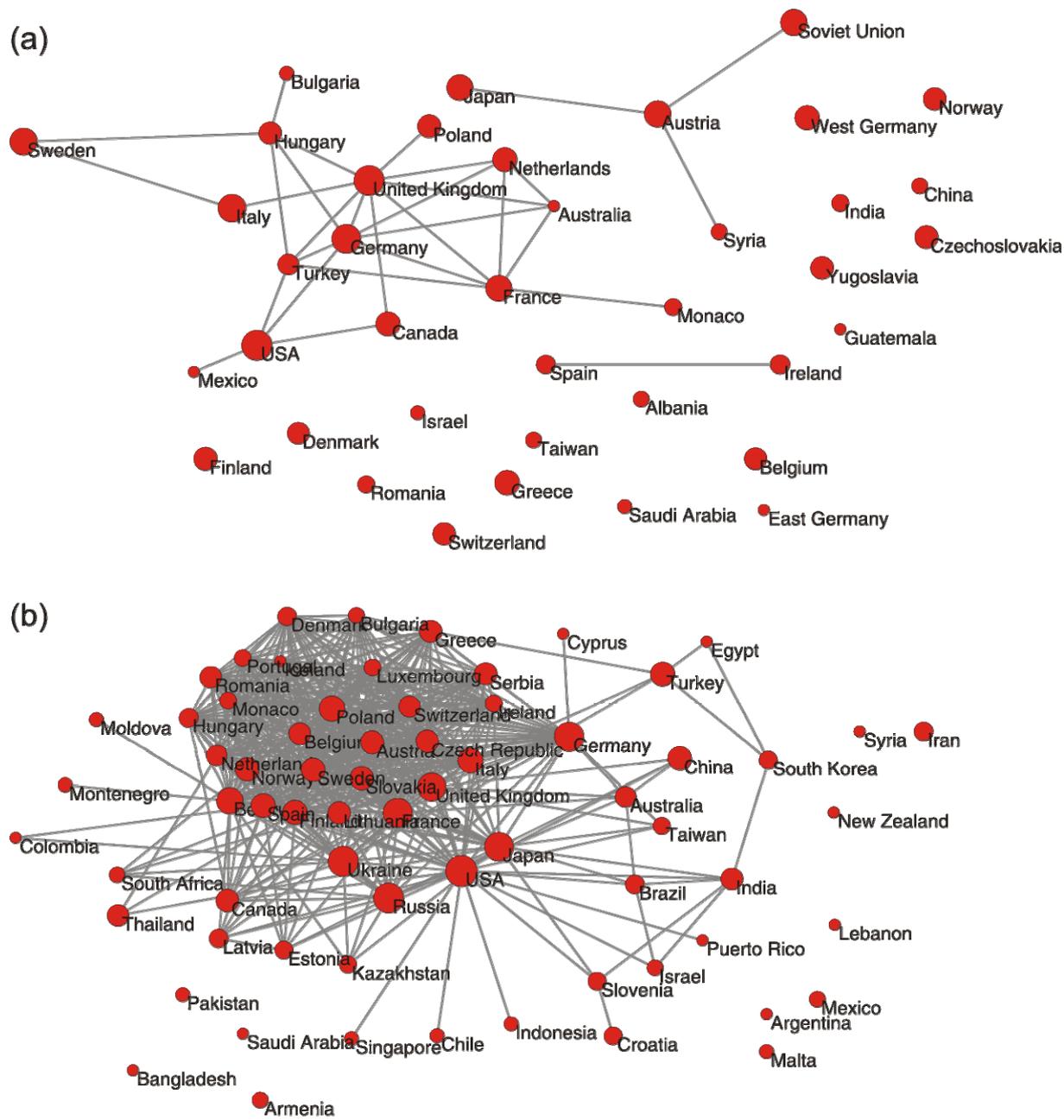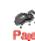

**Figure 6.** (Color online) Collaboration network of countries for Chornobyl-related research based on Scopus data. Publication period: (a) 1986–1990, (b) 2011–2015. The size of each node is proportional to the number of papers for the corresponding country.

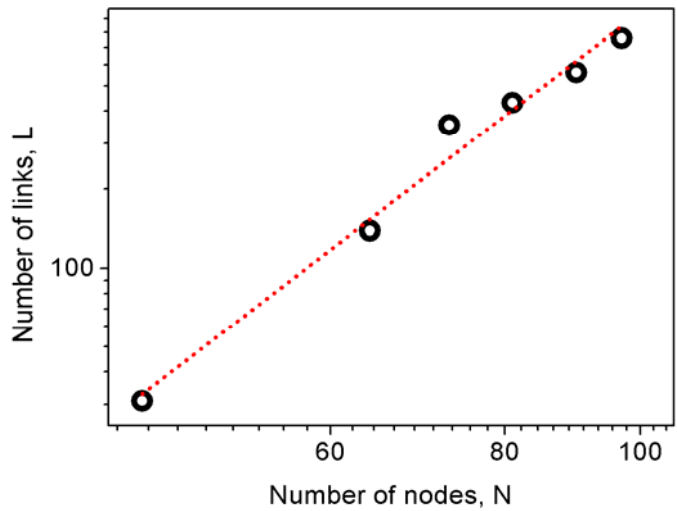

**Figure 7.** (Color online) Number of links $L$ versus number of nodes $N$ for integral network built for every 5 years starting from 1986 (symbols). The linear fit is schematically shown by broken line.

The visualization of the data can be completed by other kinds of graphics. The variation of numerical values with time is often better presented using traditional methods: e.g., the number of Chornobyl-related papers within different disciplines for different years is presented in Figure 8. The change of publication activity within TOP10 disciplines can be seen here. Two conclusions can be made instantly from the figure: there is no distinct decreasing tendency and the local maxima correspond to the "anniversary years" passed after the disaster.

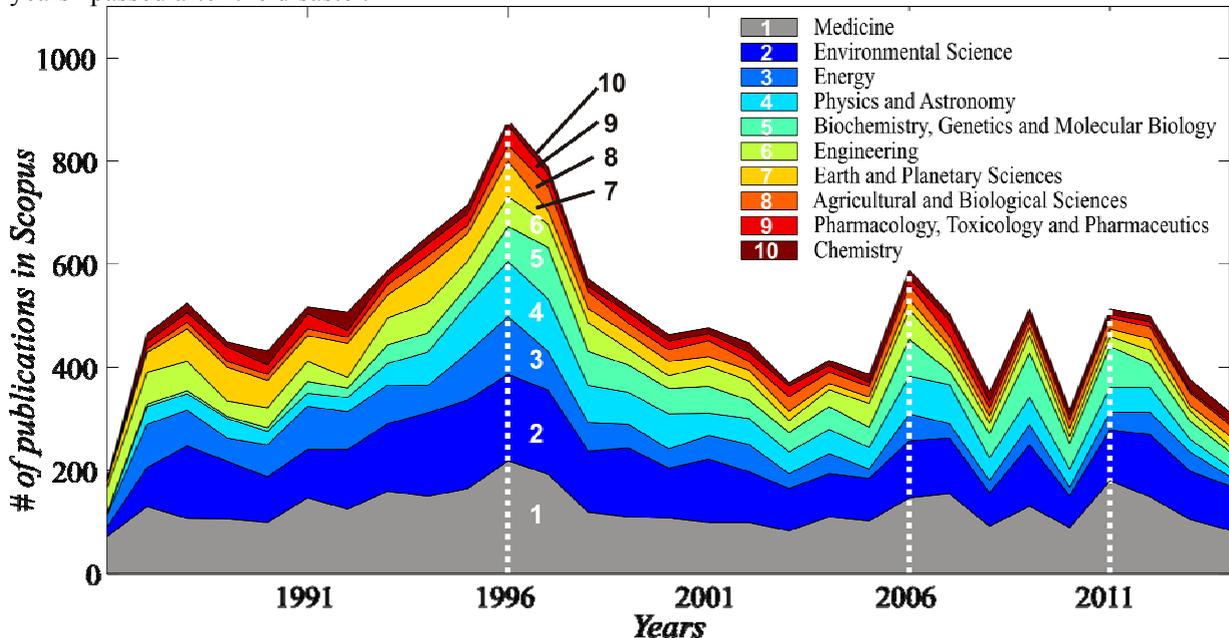

**Figure 8.** (Color online) Number of Chornobyl-related papers within different disciplines changing with time in the Scopus database, see also (Mryglod et al. 2016).

Network representation can be frustrating when the network contains a lot of links. For example, two aspects visualized in Figure 5 are also represented in Figure 9: number of countries of different part of the world and the years of their first participation. Additionally the collaboration links are shown in Figure 5, but the more connected nodes are, the less informative such visualization.

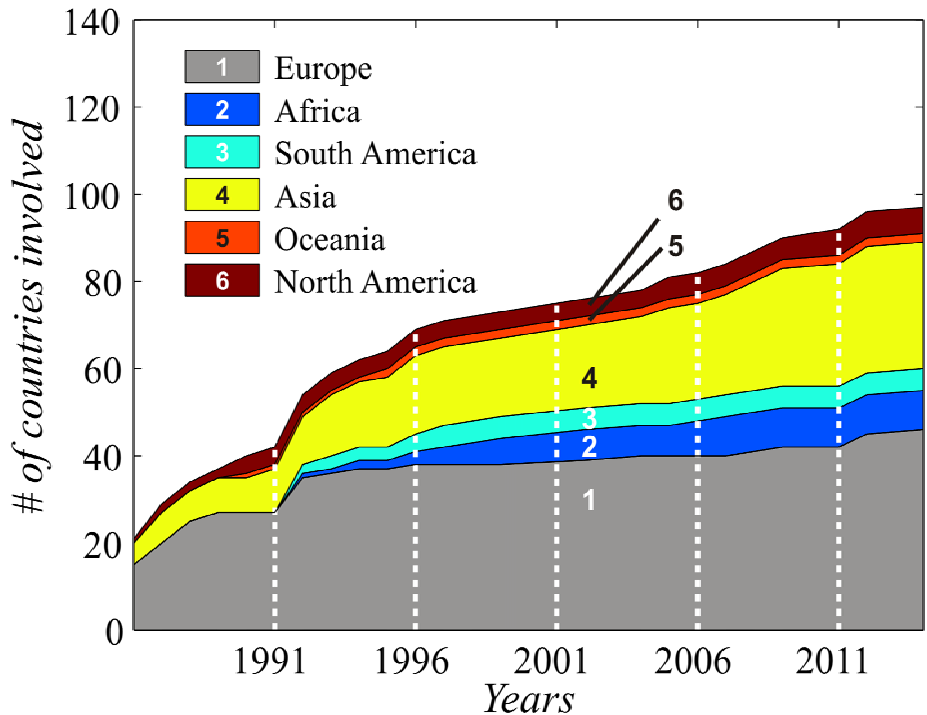

**Figure 9.** (Color online) Cumulative number of countries from different parts of the world, involved in Chornobyl-related research, see also (Mryglod et al. 2016).

## CONCLUSION

In this chapter, we have provided an example of how vizualisation of information can be used in order to deduce more insights into the problem analyzed. The example chosen is that of a of scientific publications related to the Chornobyl disaster of 1986. We have used a data basis to select relevant publications via keywords queries and we went through an analysis inspired by complex-network theory which has proven to be a very powerful tool in various other contexts, e.g. public transportation networks (Von Ferber et al 2012), mythological networks (Mac Carron and Kenna 2012). Various graphical representations were then proposed, where nodes represent the countries of authors' affiliations, and links between nodes are drawn when the countries of the nodes under interest appear in a same publication. A "neutral" vizualisation is first proposed according to the rules indicated above, which shows a giant component containing almost all nodes and a periphery of a few isolated nodes. We have then proposed other graphical representations, with e.g. the nodes with highest degrees in central positions, node sizes proportional to the number of papers, or to the node degrees, geographical characteristics of nodes of the chronology of their appearence in the set emphasized via color codes, etc. Each of these vizualisations facilitates a focus on specific features of the set analyzed.

## ACKNOWLEDGMENT

This research was supported by 7th FP, IRSES projects No. 612707 ''Dynamics of and in Complex Systems'' and No. 612669 ''Structure and Evolution of Complex Systems with Applications in Physics and Life Sciences'' and by the COST Action TD1210 ''Analyzing the dynamics of information and knowledge landscapes'' (OM and YuH) . YuH acknowledges kind hospitality of Veslava and Grzegorz Osiński at the Conference on Information visualization in humanities (Toruń, 23-24 March, 2017) where a part of this work was reported.

**KEY TERMS AND DEFINITIONS**

**Chornobyl (Chernobyl) disaster:** A catastrophic nuclear accident that occurred on 26 April 1986 in the reactor at the Chernobyl Nuclear Power Plant (Ukraine, then - USSR).

**Complex systems:** systems composed of many interacting parts, often called agents, which display collective behavior that does not follow trivially from the behaviors of the individual parts. Inherent features of complex systems incorporate self-organization, emergence of new functionalities, extreme sensitiveness to small variations in the initial conditions, governing power laws (fat-tail behaviour).

**Complex network:** a graph with non-trivial topological features that do not occur in such networks as lattices or uncorrelated random graphs. Usually it is said that any network that is more complex as classical Erdös-Rényi random graph is a complex network. Many of real-world networks manifest features of complex networks, in particular they are scale-free small worlds. Examples are given by the internet, www, technological, communication, social networks.

**Scale-free network:** a network where the probability $P(k)$ to find a node of degree $k$ (i.e. with $k$ attached links) decays at large $k$ as a power law: $P(k) \sim k^{-\lambda}$. The decay exponent plays an important role in governing different network properties.

**Scientometrics:** the field of knowledge about measuring and analysing science, technology and innovation.

**Small world network:** a network with small shortest path length (number of steps from one node to another) and high clustering coefficient (i.e. when neighbors of any node are likely to be neighbors of each other). Usually it is said that a network is a small world if its typical size $l$ grows with number of nodes $N$ slower than a power law, e.g. logarithmically: $l \sim \ln N$.